\documentclass[sigconf]{acmart}
\usepackage{array} 
\usepackage{float}
\AtBeginDocument{%
  }

\copyrightyear{2025}
\acmYear{2025}
\setcopyright{acmlicensed}\acmConference[I2M-MM '25]{Proceedings of the International Workshop on Intelligent Immersification in the Metaverse: AI-Driven Immersive Multimedia}{October 27--28, 2025}{Dublin, Ireland}
\acmBooktitle{Proceedings of the International Workshop on Intelligent Immersification in the Metaverse: AI-Driven Immersive Multimedia (I2M-MM '25), October 27--28, 2025, Dublin, Ireland}
\acmDOI{}
\acmISBN{}

\begin{document}

\title{SMARTe-VR: Student Monitoring and Adaptive Response Technology for e-Learning in Virtual Reality}

\author{Roberto Daza}
\orcid{0009-0005-2109-7782}
\email{roberto.daza@uam.es}

\affiliation{%
   \department{BiDA Lab}
  \institution{Universidad Autonoma de Madrid}
  \city{Madrid}
  \country{Spain}
}

\author{Lin Shengkai}
\email{dr.96y.8071@s.thers.ac.jp}
\orcid{0009-0009-7643-6672}
\affiliation{%
\department{Nagao Laboratory}
  \institution{Nagoya University}
  \city{Nagoya}
  \country{Japan}
}

\author{Aythami Morales}
\email{aythami.morales@uam.es}
\orcid{0000-0002-7268-4785}
\affiliation{%
   \department{BiDA Lab}
  \institution{Universidad Autonoma de Madrid}
  \city{Madrid}
  \country{Spain}
}

\author{Julian Fierrez}
\email{julian.fierrez@uam.es}
\orcid{0000-0002-6343-5656}
\affiliation{%
   \department{BiDA Lab}
  \institution{Universidad Autonoma de Madrid}
  \city{Madrid}
  \country{Spain}
}

\author{Katashi Nagao}
\email{nagao@i.nagoya-u.ac.jp}
\orcid{0000-0001-6973-7340}
\affiliation{%
\department{Nagao Laboratory}
  \institution{Nagoya University}
  \city{Nagoya}
  \country{Japan}
}

\renewcommand{\shortauthors}{Roberto Daza, Lin Shengkai, Aythami Morales, Julian Fierrez, \& Katashi Nagao}

\begin{abstract}
  This work introduces SMARTe-VR, a platform for student monitoring in an immersive virtual reality environment designed for online education. SMARTe-VR aims to collect data for adaptive learning, focusing on facial biometrics and learning metadata. The platform allows instructors to create customized learning sessions with video lectures, featuring an interface with an AutoQA system to evaluate understanding, interaction tools (for example, textbook highlighting and lecture tagging), and real-time feedback. Furthermore, we released a dataset that contains 5 research challenges with data from 10 users in VR-based TOEIC sessions. This data set, which spans more than 25 hours, includes facial features, learning metadata, 450 responses, difficulty levels of the questions, concept tags, and understanding labels. Alongside the database, we present preliminary experiments using Item Response Theory models, adapted for understanding detection using facial features. Two architectures were explored: a Temporal Convolutional Network for local features and a Multilayer Perceptron for global features.
\end{abstract}

\ccsdesc[500]{Human-centered computing~Virtual reality}
\ccsdesc[500]{Applied computing~E-learning}
\ccsdesc[500]{Computing methodologies~Machine learning}

\keywords{E-learning Platform, Virtual Reality, Facial Biometrics and Behavior, Item Response Theory, Deep Learning}
\begin{teaserfigure}
 \includegraphics[width=\textwidth]{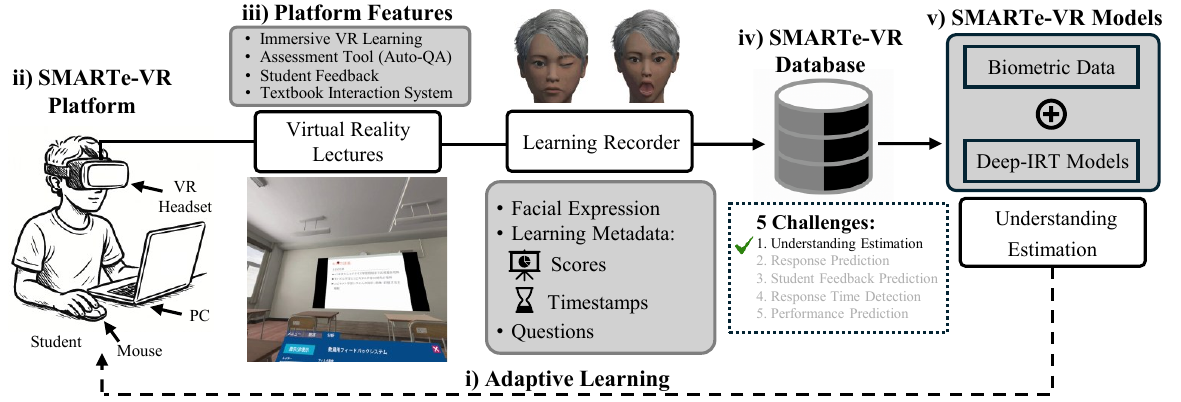}
  \caption{ Diagram of the SMARTe-VR platform for e-learning in immersive virtual reality. The figure illustrates the key features of SMARTe-VR, including its ability to capture multimodal data, such as facial expressions and learning metadata, during VR sessions to support adaptive learning. It also highlights the collected dataset, which is associated with five proposed challenges for the community, as well as the Deep IRT-based models developed to address Challenge 1: Understanding Estimation during lecture sessions using biometric data. The visual elements align with the main contributions of the paper (i–v).}
  \Description{Abstract diagram of the SMARTe-VR platform, including the database and Deep-IRT models used to estimate student understanding. The visual elements align with the main contributions of the paper (i–v).}
  \label{fig:teaser}
\end{teaserfigure}


\maketitle

\section{Introduction} \label{sec:intro}

In recent years, education has undergone a transformation driven by the digital revolution. A key factor in this change has been the growing adoption of online education or e-learning, further accelerated by the COVID-19 pandemic, which forced a transition to remote teaching models. This modality is expected to grow exponentially in the next 20 years \cite{chen2018research, daza2023edbb}.

Online education has traditionally been delivered via devices such as PC, laptop, and even tablets. However, as emerging technologies such as virtual reality (VR) have become more affordable in recent years, a question arises: Could these environments be more engaging than traditional e-learning? Through the metaverse, VR offers a sense of presence and interactivity that has often been missing in online education \cite{nagao2023virtual}.

One of VR's key strengths is its ability to more easily capture biometric and behavioral data, thanks to sensors placed on the face, hands, and other areas. This level of data collection is challenging to achieve in conventional e-learning environments. However, VR also presents challenges such as fatigue and cybersickness after prolonged use, raising questions about its viability in education ~\cite{souchet2023narrative}.

This data collection capability not only facilitates the development and adoption of new learning systems based on biometric signal analysis, behavioral data, and learning analytics, but also addresses specific challenges in online education. Among these challenges is the lack of direct contact between teachers and students, which complicates aspects such as personalized learning ~\cite{kebritchi2017issues}.

VR has the potential to improve educational quality through the implementation of Adaptive Learning (AL) systems \cite{nagao2019artificial, barbosa2024adaptive}. 
Learning analytics tools, such as Item Response Theory (IRT), play a crucial role in these systems, as they enable continuous and personalized assessment of students' knowledge and performance \cite{zhou2024state}.

The emotional and cognitive state influences their understanding and academic performance \cite{joshi2019affect,bala2023implementation}. For this reason, recent research has expanded the scope of Learning Analytics (LA) to Multimodal Learning Analytics (MMLA).
Powered by advanced technologies such as deep learning, these approaches have achieved state-of-the-art results in predicting academic performance or knowledge. In particular, facial biometric signals have shown a clear relationship with factors such as attention, stress, and emotions, highlighting their potential to optimize adaptive and personalized learning \cite{hernandez2019edbb, daza2023edbb, kovanovic2023unobtrusive}. VR offers a significant advantage in capturing facial expressions more easily and in an environment perceived as more realistic by the user, through VR headsets with facial sensors that allow detailed and continuous tracking of user interaction.

However, current public databases in e-learning have several limitations. 
\begin{itemize} 

\item To our knowledge, there are no public databases specifically focused on VR environments for e-learning, as most are centered on traditional e-learning environments.

\item Most databases are not multimodal; they do not include biometric or behavioral information. 

\item The majority focus solely on learning assessment by collecting only responses to problems or questionnaires, without considering the continuous monitoring of the student's Learning Session (LS). 
\end{itemize}

These limitations hinder the study of adaptive learning (AL) systems in VR, particularly for data-driven multimodal models.  They also prevent comprehensive comparisons of the advantages and drawbacks relative to traditional e-learning environments.



The contributions of this work are shown in Fig. \ref{fig:teaser} and detailed as follows: i)  We present a survey of state-of-the-art AL and public e-learning databases. ii) We develop a visualization platform for VR, integrating an AutoQA system to assess student comprehension. This platform enables continuous monitoring by capturing metadata and facial expressions. iii) We designed an online session protocol, which includes an initial video lecture paired with a pretest, 10 video lectures with accompanying questions of varying difficulty levels, along with an automated note system using ChatGPT. This system allows students to quickly highlight information in VR environments using a color-coding scheme, facilitating review. iv) We have released an initial subset of the database (SMARTe-VR-db\footnote{Available at: \url{https://github.com/BLANCELIN1/SMARTe-VR-DB}}), featuring 10 VR users. Each Learning Session (LS) lasts 2.5 hours and captures metadata and facial features. The database was carefully designed to create a balanced dataset, specifically designed to train and evaluate data-driven interpretability models for AL. And v)~ we conducted preliminary experiments based on Item Response Theory (IRT) systems to validate the data set, using local and global facial features collected during lecture videos to predict whether the student has understood the content.

The remainder of the paper is organized as follows. Section~\ref{s:Related Works} provides a summary of related work on AL with a focus on IRT and public databases in e-learning. Section~\ref{s:Platform} presents the VR visualization platform, the published database, and the evaluation challenges proposed for the community. Section~\ref{s:Experiments}  outlines the experiments and results. Finally, conclusions and directions for future work are discussed in Section~\ref{s:Conclusion}.

\section{Related Work: Adaptive Learning} \label{s:Related Works}

Adaptive Learning (AL) is an educational approach that personalizes the learning experience for each student by dynamically adjusting content, pace, and learning pathways according to individual needs and responses \cite{barbosa2024adaptive}. This is relevant in e-learning, where students often receive identical resources without considering prior knowledge or skills, and direct interaction is limited \cite{mccusker2013intelligent}.
AL has gained momentum in recent years, driven by advances in Artificial Intelligence (AI), Machine Learning, and data analytics. 

\subsection{Adaptive Learning: Approaches} 

Traditionally, learning style models, such as Felder-Silverman and Kolb \cite{felder2002learning, kolb2007kolb}, have been used to personalize teaching; however, their effectiveness has been questioned due to the lack of solid empirical evidence. These models classify students according to sensory and cognitive preferences \cite{an2017learning}. In response, modern adaptive systems prioritize objective data analysis, such as performance analytics and biometrics. 


\begin{table*}[h!]
\caption{Summary of existing datasets for research in adaptive learning (AL) \cite{assistments2012, chang2015modeling, choi2020ednet, zhou2024state, hernandez2019edbb, daza2024improveimpactmobilephones} in comparison to our contributed SMARTe-VR-db. Bio. $=$ Biometrics, Eval. $=$ Evaluation, LS $=$ Learning Session, and Logs $=$ Recorded  material consulted by students.}
\centering
\begin{tabular}{|>{\centering\arraybackslash}m{2.6cm}|>{\centering\arraybackslash}m{1.1cm}|>{\centering\arraybackslash}m{0.6cm}|>{\centering\arraybackslash}m{0.6cm}|>{\centering\arraybackslash}m{2cm}|>{\centering\arraybackslash}m{8cm}|}
\hline
\textbf{Dataset} & \textbf{Users} & \textbf{VR} & \textbf{Bio.} & \textbf{LS} & \textbf{Description} \\
\hline
ASSISTments \cite{assistments2012}& 46,674 & No & No & Eval. only & Math responses (1 year); thousands of questions \\
\hline
Junyi \cite{chang2015modeling} & 247,606 & No & No & Eval. only & Math responses; 722 questions \\
\hline
EdNet (KT4) \cite{choi2020ednet} & 297,915 & No & No & Logs \& Eval. & TOEIC exam responses with logs of consulted material (15 months); thousands of questions \\
\hline
SAD-IRT \cite{zhou2024state} & 20 & No & Yes & Eval. only & 17 hours of assessments with facial tracking; 1,000 multiple-choice responses \\
\hline
edBB \cite{hernandez2019edbb} & 60 & No & Yes & Eval. only & 10 hrs assessment, focused on behavioral detection \\
\hline
IMPROVE \cite{daza2024improveimpactmobilephones} & 120 & No & Yes & Class \& Eval. & Primarily focused on effects of phone use on learning\\
\hline
\bf{SMARTe-VR-db (Ours)} & 10 & Yes & Yes & Class \& Eval. & 25 hours  in VR, covering learning and assessment with facial tracking; 450 multiple-choice responses \\ 
\hline
\end{tabular}

\label{table:dataset_summary}
\end{table*}

Modern adaptive education systems aim to model an abstract representation of a student's skill or level of knowledge based on their performance and interactions with the educational system. This is a continuous process, as the students evolve over time \cite{desmarais2012review}. The technological foundation lies in LA, an area that collects and analyzes data on student interactions and engagement, providing real-time insights for AL adjustments \cite{chatti2012reference}. The field is advancing towards MMLA, integrating diverse data sources such as biometrics, behaviors, and metadata to offer a more holistic view of the educational process \cite{hernandez2019edbb, daza2023edbb}. Three student models have been widely developed:
\begin{itemize}
\item Item Response Theory (IRT):   A statistical model designed to evaluate both student ability (\( \theta \)) and question difficulty (\( b \)) based on their responses, making it highly suitable for creating adaptive tests. IRT models utilize estimation methods such as Maximum Likelihood Estimation (MLE) or Bayesian inference, often employing iterative techniques for parameter estimation. IRT models vary according to the parameters they incorporate. The most common version is the 1PL (Rasch model), which includes only the difficulty parameter (\( b \)) for each item \cite{baker2001basics, himelfarb2019primer}.

    
    

Since traditional IRT models typically assume that examinees' abilities are randomly sampled from a standard distribution, recent research has explored combining IRT with deep learning techniques. A notable example is the Deep-IRT model \cite{tsutsumi2021deep}, which introduces two twin neural networks: one to model the ability of the student and the other to model the difficulty of the elements. The network inputs are one-hot vectors for the question and the student. This model outperforms traditional models across diverse datasets, predicting unknown item responses more accurately from previous histories, and proving effective even for small datasets. However, it may encounter challenges with sparse data.

Recent studies have adapted the 1PL model with deep learning to incorporate biometric signals, such as facial expressions, to predict an additional parameter: the student's state. One prominent example is the SAD-IRT model \cite{zhou2024state}, which adjusts item difficulty dynamically based on students' cognitive and emotional states. Based on Deep-IRT, SAD-IRT adds a branch with spatial extractors (MARLIN) \cite{cai2023marlin} and temporal features (Temporal Convolutional Network, TCN). This branch processes facial images captured while students solve test items to derive the student state parameter.  Evaluated on a dataset of 20 participants and 1,000 responses, SAD-IRT outperformed both the DEEP-IRT model and the facial video-only models.

\item Knowledge Tracing (KT): Models the evolution of a student's knowledge by updating an internal "knowledge state" based on responses over a sequence of questions, using techniques like the Hidden Markov Model. Unlike IRT, KT focuses on sequential learning rather than static skills, without accounting for the difficulty of the question or individual ability. Advanced KT models, such as Deep KT, leverage deep learning for improved response prediction accuracy, although they lack the interpretability of IRT \cite{zhou2024state}. Combined IRT-KT methods have shown promising results, in addition to recent efforts to integrate biometric and behavioral data to capture affective states \cite{ghosh2020context}.
\sloppypar
\item Biometrics- and Behaviour-Based Learning Recognition (BBLR): These emerging models assess the student's level of understanding during the session, without requiring constant feedback through answers. They integrate biometric and behavioral data, such as facial expressions \cite{zhou2024state}, eye blinks \cite{daza2020mebal, daza2024mebal2}, visual attention \cite{Navarro2024}, keystroke dynamics \cite{morales2015biometric, morales2016kboc}, heart rate \cite{hernandez2020heart}, head pose \cite{daza2024improveimpactmobilephones, Becerra2024}, etc., to estimate cognitive load \cite{daza2021alebk, daza2023matt, daza2024deepface}, emotions \cite{bala2023implementation}, etc., linking these states to learning levels. Unlike traditional approaches such as IRT and KT, these models enable real-time feedback, adjusting content, and pacing according to the student’s state. Their main challenge lies in a lack of interpretability due to complex deep learning architectures, which do not always clearly interpret the factors that contribute to academic performance. 
\end{itemize}


\subsection{Adaptive Learning: Platforms} 

Several educational platforms have adopted AL models. For example, Duolingo uses IRT models to optimize language learning \cite{maris2020duolingo}, and platforms like ASSISTments and Coursera employ KT models or similar methods \cite{feng2009addressing}. Some public institutions have also developed AL systems, e.g.:


\begin{itemize}

\item Learning Evidence Analytics Framework (LEAF): Developed at Kyoto University \cite{ogata2022learning}, LEAF applies learning analytics to collect, analyze, and visualize educational data. Includes tools such as BookRoll, Log Pallet, and LAView for real-time data visualization and feedback. Through AI-driven big data analysis, LEAF provides personalized learning support, enabling immediate feedback for students and instructors.


\item edBB: Developed at the Autonomous University of Madrid \cite{hernandez2019edbb, daza2023edbb}, edBB focuses on learning recognition through biometric and behavioral analysis, using CNN and RNN models to estimate attention, heart rate, and more. Integrated with M2LADS \cite{becerra2023m2lads, becerra2023user, BecerraM2LADSDEMO2025}, it offers instructors a comprehensive view of student progress. 


\end{itemize}

\subsection{Existing Research Datasets}

Public data sets in education focus mainly on student responses and assessment (see Table \ref{table:dataset_summary}), with limited incorporation of biometric data and without VR. For example, the ASSISTments \cite{assistments2012} and Junyi \cite{chang2015modeling} data sets provide responses to math questions. Similarly, the EdNet dataset \cite{choi2020ednet} includes question responses and incorporates records of materials consulted by students during the preparation for the TOEIC exam. All of these data sets lack biometric and VR monitoring.

Few datasets explore multimodality with biometrics. The SAD-IRT dataset \cite{zhou2024state} includes facial tracking during literacy tasks, while edBB \cite{hernandez2019edbb} records biometric signals such as heart rate, EEG, and facial video for behavior analysis. Similarly, the IMPROVE \cite{daza2024improveimpactmobilephones} dataset captures multimodal biometric and behavioral data during short e-learning sessions, with assessment being the main focus of data collection. IMPROVE was specifically designed to investigate the effects of mobile phone usage on student learning \cite{becerra2025multimodal}. However, these data sets focus on PC-based assessment.

In contrast, SMARTe-VR uniquely combines VR, learning, and assessment with biometric monitoring, making it a valuable resource for the development of AL systems.

\section{SMARTe-VR: Platform Description}  \label{s:Platform}

\begin{table*}[h!]
\caption{Overview of devices and data collected by the monitoring system.}
\centering
\begin{tabular}{|>{\centering\arraybackslash}m{4cm}|m{4cm}|m{8.5cm}|}
\hline
\textbf{Device} & \textbf{Features} & \textbf{Information} \\
\hline
VIVE XR Elite & 90Hz \newline 1920x1920 resolution \newline $110^\circ$ field of view \newline 4 tracking cameras \newline Speakers & Visual display and audio, spatial tracking, and motion detection, responsible for transmitting data to the PC. \\ 
\hline
VIVE XR Facial Tracker & 30Hz \newline Dual-tracking eye camera \newline Mono-tracking face camera & Attachable to the VR headset, tracks facial expressions and movements, and detects interpupillary distance. \\
\hline
PC & Mouse used as a controller \newline \raggedright Data processing & Records: 51 facial features, timestamps, interaction buttons, video feedback tags, textbook interactions (e.g., consulted pages), lecture videos, questions, responses,  selection history, scores, understanding labels, pretest results, content/question difficulty, and question concept tags. \\
\hline
\end{tabular}
\label{table:vr_setup}
\end{table*}

SMARTe-VR platform is designed for VR-based e-learning sessions, allowing instructors to upload their educational content while students experience it in an immersive environment that simulates a lecture hall with a screen, similar to works such as \cite{nagao2023virtual}. The platform provides a virtual reality environment that facilitates student interaction with the system, monitors learning data, and captures facial expressions in real time. This feature allows for a detailed analysis of the student's emotional and cognitive state. In addition, it integrates learning assessment tools, allowing instructors to adjust content based on student performance.

\subsection{Platform Features}


\noindent \textbf{VR Interaction:} The immersive environment simulates a
virtual lecture hall, allowing students to zoom in/out by
moving their head, adjust brightness, and interact using VR
controllers or a mouse. Features include answering questions, opening the textbook, highlighting text, and controlling
video playback (pause, fast-forward, ~etc.).

\noindent \textbf{Structure and Assessment Tool:} Instructors can structure LSs using lecture videos of various lengths and content. It also features an automatic assessment system (Auto QA) that can be activated at different stages. \textit{Before the session:} Preliminary questions assess initial student knowledge. \textit{After each video:} Comprehension of the content is assessed. This method, used by platforms such as Coursera and edX, is supported by research showing that a few well-crafted questions effectively measure real-time understanding \cite{cummins2015investigating}. These assessments support models such as IRT and KT, enabling the platform to be used for intelligent AL.

\noindent \textbf{Student Feedback:} After each Auto QA intervention, students receive their results and a comparison with the highest performing student. At the end of the session, a summary of the scores is provided. Continuous feedback and performance comparisons motivate students, helping them track progress, identify focus areas, and optimize learning using notes and resources \cite{van2015effects}.

\noindent \textbf{Textbook Interaction System:} SMARTe-VR generates summaries of main ideas from each lecture video using subtitles and a Large Language Model (ChatGPT). Instructors can review and edit these summaries to create a textbook. Students can view and highlight the text anytime, even during Auto QA questioning.
To encourage interaction and collect learning data, two methods are used:
\begin{itemize}

\item Textbook Highlighting: Students can mark text using a four-color coding system, where each color represents a distinct function: learned content, material for review, lack of understanding, and key content for memorization.

\item Quick Tagging Buttons: A set of four color-coded buttons allows students to quickly tag video sections, using the same meaning as in text highlighting. The platform labels these sections, providing valuable information for instructors and learning analytics.
\end{itemize}
Studies show that visual interaction mechanisms improve content retention and comprehension. Highlighting key information is associated with higher retention and more efficient review \cite{ogata2022learning}.



\makeatletter
\renewcommand{\@makefntext}[1]{\noindent\makebox[0pt][l]{\hspace{0.5em}\textsuperscript{\@thefnmark}}#1}
\makeatother

\subsection{Contributed Database: SMARTe-VR-db}

We monitor LSs focused on the TOEIC, specifically Section 5, which assesses grammar comprehension. The TOEIC is a recognized English test used by companies to evaluate nonnative speakers' language skills in professional contexts. This study involved 10 Japanese students from Hiroshima University.
The data set comprises 450 answered questions and approximately 25 hours of learning data, including facial movements and expressions. The SMARTe-VR-db is publicly available on GitHub \cite{daza2025smartevrdb}.


\noindent \textbf{Ethical Considerations:} The database was collected in accordance with the Declaration of Helsinki and approved by the Ethics Committee. The learners were briefed about the procedure and data acquisition, providing informed consent. Due to extended monitoring, participants received an incentive with an additional reward for top performers.


\subsubsection{Devices} 

Table \ref{table:vr_setup} lists the devices and data types captured. The setup included the following components (see Fig. \ref{fig:Diagram}):

\noindent \textbf{VR Headset:} We used the VIVE XR Elite, designed for educational and business applications, together with the VIVE XR Facial Tracker. This accessory uses infrared cameras to measure interpupillary distance for optimal viewing and facial tracking. Using the OpenXR SDK, we monitored up to 51 facial features at 30 Hz, ensuring system stability and preventing overheating. Captured facial features include movements and expressions, quantified from 0 to 1 intensity. These features include eye movements and blinking, jaw opening, mouth and lip expressions (e.g. smiles), cheek gestures (e.g. puffing) and tongue movements. The VIVE XR Facial Tracker combines dual infrared cameras for eye movement tracking (120 Hz) and a monochrome camera (60 Hz) with an ultra-wide 151° field of view to capture facial movements below the eyes, enabling reliable detection of subtle expressions even under low-light conditions. The combination of a wide-angle field of view, inertial sensors, and facial cameras embedded in the VR headset ensures stable tracking during head motion, full rotations, or suboptimal device alignment (typical limitations of standard webcams used in traditional e-learning setups). The immersive VR context further promotes natural and expressive facial behavior, improving the fidelity of facial biometrics \cite{dubovi2024facial}.

\noindent \textbf{Personal Computer:} Connected to the VIVE XR Elite, it managed the virtual environment, collected and analyzed sensor data, and recorded learning logs while maintaining optimal performance without overheating the headset. Although VIVE XR Elite controllers are supported, we chose the mouse as the VR controller because: 1) students remained seated, simulating a classroom, making it more practical; 2) test feedback indicated it was more comfortable to interact with the SMARTe-VR interface and highlight text; 3) this approach allows for better comparisons to traditional e-learning setups for future research.


\subsubsection{Protocol and Tasks} 
The goal was to design a monitoring system for real VR LSs, covering both learning and assessment processes while tracking learning metadata and facial expressions. This protocol aims to help the research community in developing frameworks like IRT, KT, MMLA, and BBLR, filling the gap in research data sets with these features.

The content of each lecture video and its questions was designed to create a balanced dataset, with an equal number of videos labeled understood or not understood. Ten 8-minute TOEIC exam lecture videos, explained in Japanese, were selected from YouTube and varied in difficulty according to the TOEIC guidelines. Each video had three multiple-choice questions with four options, focused on sentence completion to assess grammar and vocabulary. A video was labeled understood if a student answered at least 2 out of 3 questions correctly. The questions were from a TOEIC data set on the ProProfs platform, which includes data on the precision of the response and the number of participants.

Using sample size estimation, we determined the minimum number of responses needed to represent 99\% of the population with a 1\% margin of error, allowing the classification of questions by difficulty and content. The questions were classified as easy (80\% correct response rate), medium (50\%), and difficult (20\%). The question structure per video was the following: easy-level videos had 2 easy and 1 medium question; medium-level had 3 medium questions; difficult-level had 2 difficult and 1 medium question. To verify difficulty, a preliminary test was conducted with similar students, collecting subjective evaluations on the difficulty of each session and its questions.

Each student participated in a 120-150 minute LS consisting of 11 lecture videos and 45 questions, similar to a university class. The sessions included:

\begin{figure*}[t]
    \centering
    \includegraphics[width=\textwidth]{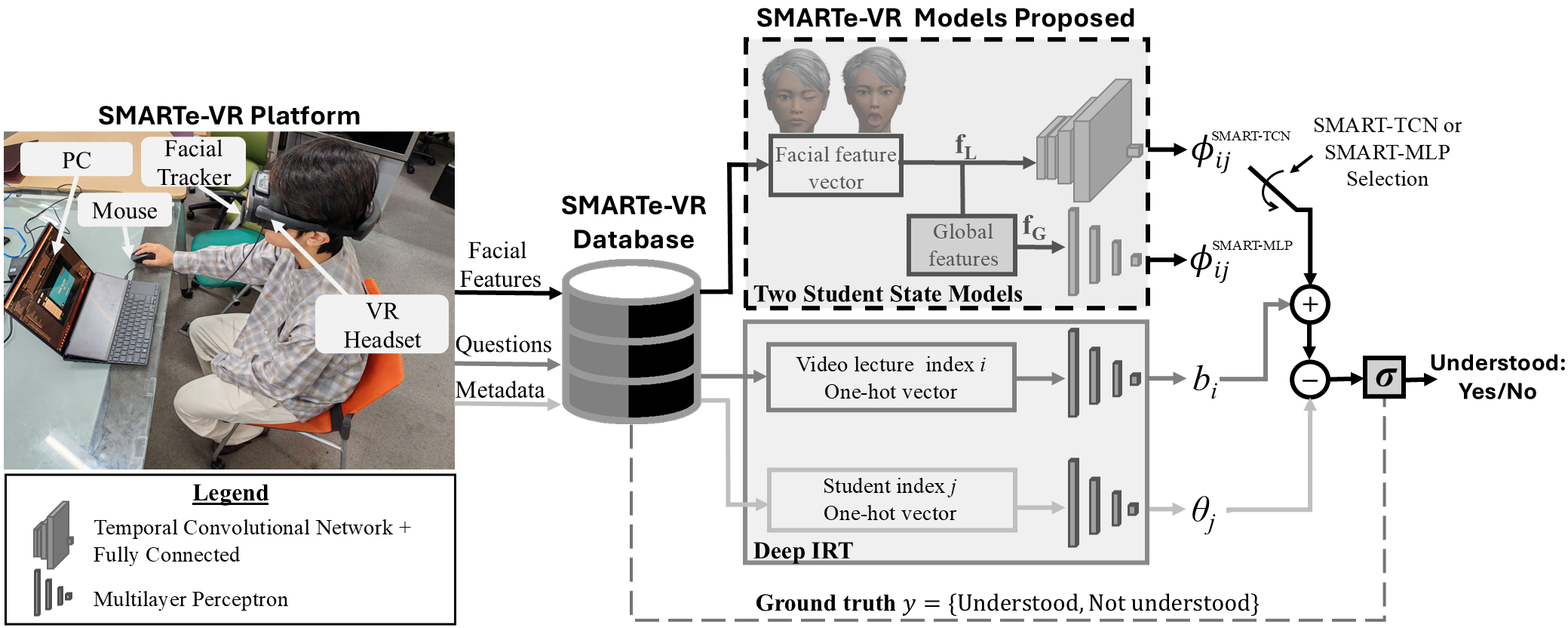} 
    \caption{Example setup and data captured by the SMARTe-VR platform, including the proposed understanding detection models (SMART-TCN and SMART-MLP) for database validation. Both follow a Deep IRT-based architecture \cite{tsutsumi2021deep}  for estimating student ability and item difficulty from one-hot encodings of users and questions. They also use facial features to estimate the student state, which is their only point of difference: SMART-TCN applies a TCN to local facial features, while SMART-MLP uses an MLP on global facial features.}
    \label{fig:Diagram}
    \Description{The figure shows the setup of the SMARTe-VR platform, the data captured during the learning sessions, and how this information is used by two proposed IRT-based models—SMART-TCN and SMART-MLP—to predict student understanding during video lectures.}
\end{figure*}

\noindent\textbf{Before Starting the Session:} \textit{Calibration and Student Information:} The supervisor explained the protocol and verified data capture. A 3D facial model was monitored to confirm the recording of facial movements and expressions. \textit{Privacy Policy:} The students reviewed a document detailing data collection and its purpose, with the option to accept or decline participation. \textit{Video Lecture 0 and Auto QA Trial Phase:} The students began with an initial lecture video explaining how to use the interface. At the end of this video, the Auto QA system was activated, presenting 5 questions without a time limit to help students become familiar with the interface. \textit{Auto QA Pretest:} A pre-test of 10 questions (3 easy, 4 medium, and 3 difficult) related to the upcoming lecture content was designed with two main objectives: \textit{i)} to evaluate the current level of knowledge of the students and \textit{ ii)} to serve as a baseline for IRT or similar systems.

\noindent\textbf{Learning Session:} The session consisted of 10 lecture videos, each followed by 3 related questions. The structure was as follows: i) Students watched an 8-minute video, interacting with the textbook and color-coded buttons. ii) Auto-QA was activated after the video, presenting 3 questions with a one-minute response time each, allowing students to review the textbook. This immediate questioning reduced potential memory problems. iii) Students received feedback and compared their performance to the top student. iv) Students could take a break if needed; most did not, but those who did paused for approximately 5 minutes. And v) a 5--10 minute battery change of the VR headset occurred after the fifth video.

\noindent\textbf{End of the Session:} The students received a total score and a comparison with the highest performing student. They provided feedback to the supervisor about their SMARTe-VR experience, which highlighted the following:
i) prolonged VR use did not cause motion sickness as participants remained seated in the VR, ii) students reported restlessness over time due to limited interaction options, and iii) some students reported discomfort when using the mouse in VR, as they were unable to see their hands. We conclude that integrating Mixed Reality (MR) and shortening sessions could enhance interaction, device usability (e.g. mouse, keyboard) and overall user experience.



\subsubsection{Challenges}

We propose 5 challenges for the research community, each with a specific target and input data. The objective is to take advantage of the SMARTe-VR database to develop AL models that estimate student performance, knowledge, or behavior. The proposed challenges are as follows. 

\noindent \textbf{Challenge 1 - Understanding Estimation:}  Predicting student comprehension of educational content is valuable for AL. We propose estimating whether a student understood the lecture content using binary labels (understood, not understood). \textbf{Target:} Binary classification to predict student understanding in VR based on multimodal data. \textbf{Inputs:} Facial features, metadata (e.g. highlighting, button presses), pre-test results and content difficulty.

\noindent \textbf{Challenge 2 - Response Prediction:} Predicting student responses using IRT or KT models with multimodal data. Proposed scenarios: i) response data only, ii) adding multimodal data captured during the response (one minute / question), and iii) all data from content and responses. \textbf{Target:} Predict correct answers. \textbf{Inputs:}  450 questions, facial features, metadata (e.g., response time), question concept tags, content and question difficulty, and pretest results.

\noindent \textbf{Challenge 3 - Student Feedback Prediction:} The ability to anticipate how a student perceives content (e.g., important, learned/not learned, to memorize) is crucial for AL models. This challenge seeks to create a model that uses color-coded button interactions from VR sessions as ground truth to predict feedback.  \textbf{Target:} Predict student feedback on their understanding and perceived relevance of the content. \textbf{Inputs:} Facial features, learning metadata, pretest results, and lecture videos. 

\noindent \textbf{Challenge 4 - Response Time Detection:} Predicting if a student will respond before the time limit can help gauge confidence, detect fatigue or inattention, and dynamically adjust workload or response time. Only biometric data and metadata from the first 20 seconds are allowed. \textbf{Target:} Predict if a student will respond on time. \textbf{Inputs:}~ Facial features,  metadata (e.g., selection path history), content and question difficulty, and pretest results.

\noindent \textbf{Challenge 5 - Performance Prediction:} Student performance was measured by accuracy (percentage of correct responses), allowing classification into three levels (low, medium, high). We propose estimating the final performance across three levels using biometric data and session metadata without incorporating response data. Indicators such as attention, emotional state, and fatigue of facial features, along with learning metadata, can help predict performance. \textbf{Target:} Performance. \textbf{Inputs:} Facial features, learning metadata, content difficulty, and pretest results.

\section{Experiments and Results} \label{s:Experiments}

\subsection{IRT Models}
We tested two IRT (1PL) models based on \cite{zhou2024state} to validate our methods for Challenge 1: to predict student understanding during video lectures. These models use facial features in a time window ${W_l \in \{1,\ldots,8\}}$ (in minutes), along with one-hot vectors for the user and the questions. The binary understand label served as the ground truth for prediction. Fig. \ref{fig:Diagram} shows a schematic of these ~models:

\noindent \textbf{SMART-TCN model:} Based on Deep-IRT (see Section~\ref{s:Related Works}), we used two twin Multilayer Perceptrons (MLPs) to estimate the difficulty of the question (\( b \)) and the ability of the student (\( \theta \)) separately. An MLP processed a one-hot vector representing the question to obtain \( b \), while the other processed a one-hot vector identifying the student to estimate \( \theta \). The student state (\( \phi \)) was based on SAD-IRT, using a TCN that processed facial features extracted from video lectures using the VR headset. The TCN was configured with 4 layers (32 to 64 neurons), a filter size of 3, and an expansion rate of 1, 1, 2, and 4, optimized to capture subtle changes in expressions held for extended periods. The TCN output was then connected to a fully connected layer with an output layer using a Tanh function to obtain the state parameter. Finally, the system estimated the probability of understanding as: 
\begin{equation}
P(\text{Understand}) = \frac{1}{1 + e^{-(\theta - (b + \phi))}}
\end{equation}



\noindent \textbf{SMART-MLP Model:} This model was similar to SMART-TCN but replaced the TCN with an MLP for student state estimation, resulting in a structure with three parallel MLPs to estimate each parameter independently. Each MLP had 2 hidden layers, the first with 128 neurons and the second with 64 neurons, both using ReLU activation, followed by an output layer with a Tanh function. This model used global facial features as input to estimate \( \phi \).

Both models were trained using the Adam optimizer with a learning rate of 1e-03, a batch size of 64, and a total of 200 epochs. A cross-entropy loss function was applied for binary prediction of understanding.

\noindent \textbf{Local Features.} For the SMART-TCN model, we used local facial features, with the VR headset providing 30 sets of 51 features per second. The resulting vector is $\textbf{f}_{\mathrm{L}} \in \mathbb{R}^{30 \times 51 \times W_{l} \times 60}$, with \( W_l \) being the time window in minutes.

\noindent \textbf{Global Features.} For the SMART-MLP model, we used 8 global statistics for each facial feature. These are the maximum, minimum, mean, standard deviation, median, kurtosis, skewness, and rate of change. The resulting vector is $\textbf{f}_{\mathrm{G}} \in \mathbb{R}^{8 \times 51}$.

\noindent \textbf{Traditional IRT baseline.} To compare the proposed models, we implemented a traditional IRT model (1PL, Rasch) as a baseline, using the following estimation process. First, item difficulties were approximated using a sign-inverted Probit transformation of the empirical accuracy rates as initial values. Then, the abilities of the students were estimated by MLE by maximizing the likelihood of their observed responses. Both item difficulties and student abilities were updated in an iterative process until convergence \cite{baker2001basics}.

\begin{figure}[t]
    \centering
    \includegraphics[width=\columnwidth]{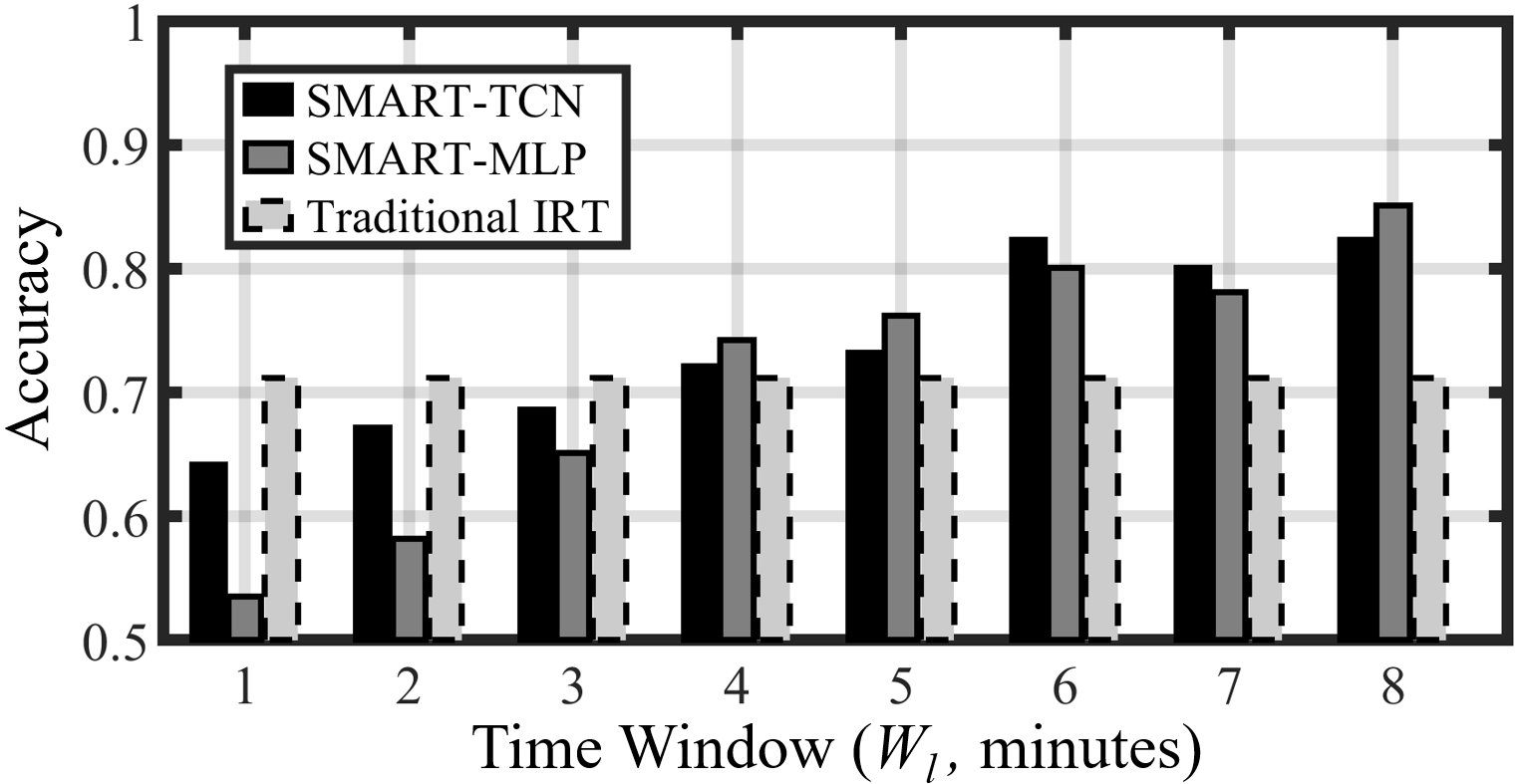} 
    \caption{Comparison of accuracy across different time windows (\( W_l \)) for our proposed models and the Traditional IRT  baseline on the SMARTe-VR database.}
    \Description{Bar chart showing the accuracy obtained on the SMARTe-VR database for three models: SMART-TCN, SMART-MLP, and the Traditional IRT baseline.}
    \label{fig:Dist_Y}
\end{figure}

\subsection{Experimental Protocol}

SMARTe-VR is a nearly balanced data set on understanding labels, with 55\% of sessions marked as understood. Specifically, 83\% of easy, 45\% of medium, and 40\% of difficult sessions are marked as understood.  Each user has 10 labels for each learning session, totaling 100 labels for all users.

We used a leave-one-out cross-validation approach, where each user is left out once for testing while the remaining users are used for training; this process repeats for each user. However, even with leave-one-out, the model is always trained on video lecture 0 and the 10 pretest questions, including those from the left-out user, to estimate their initial ability values. The decision threshold is set at the point where the False Positive and False Negative rates are equal.

\subsection{Results}

Fig.~\ref{fig:Dist_Y} presents the results obtained for the proposed models compared to traditional IRT for various time windows ${W_l \in \{1,\ldots,8\}}$ (in minutes).

The results indicated that with short time windows, the traditional IRT outperformed the two deep learning-based models, which struggled in this range. However, from 4 minutes onward, facial features began to provide valuable insights, enhancing both models’ ability to predict student state and understanding beyond the capabilities of traditional IRT. Both models showed better performance as \( W_l \) increased, suggesting that facial features more accurately represented the student's state over longer periods. With time windows of several minutes, the proposed models were better able to capture behavioral patterns in facial expressions.
Furthermore, given that Japanese students tend to maintain more neutral and consistent facial expressions compared to other cultures, facial changes were likely subtler and less frequent. This suggests that, with shorter time windows, the collected facial signals may not have adequately reflected the student's comprehension state.
SMART-MLP, based on global features, achieved the highest accuracy of 85\% in the 8-minute window. Both models demonstrated similar performance across all time windows, though SMART-MLP performed slightly worse in shorter windows, likely due to the limitations of global features in capturing rapid patterns. 

Finally, for the SMART-MLP model in the 8-minute time window, which achieved the best results, we analyzed its accuracy by the content difficulty levels provided in the dataset (see Table \ref{table:performance_difficulty}). The SMART-MLP model outperformed the traditional IRT model in all content. Both models showed a similar trend, being more effective in detecting understanding for medium-level content, followed by difficult-level content, with the lowest accuracy for easy-level content. With more medium-difficulty sessions in the dataset, SMART-MLP can more easily identify patterns. Additionally, IRT models benefit from medium-difficulty items, as the variability in responses allows the model to better discriminate between skill levels, crucial for accurate estimates.

\begin{table}[t!]
\caption{Accuracy obtained for the proposed SMART-MLP model with 8-minutes time windows compared to the traditional IRT baseline, evaluated across different content difficulty levels in our contributed SMARTe-VR-db.}
\centering
\renewcommand{\arraystretch}{1.5} 
\begin{tabular}{|l|c|c|c|}
\cline{2-4}
\multicolumn{1}{c|}{} & \multicolumn{3}{c|}{\textbf{Content Difficulty}} \\
\hline
\textbf{Model} & \textbf{Easy} & \textbf{Medium} & \textbf{Hard} \\
\hline
SMART-MLP  (\( W_l = 8 \))     & 0.828 & 0.868 & 0.851 \\
\hline
Traditional IRT & 0.682 & 0.741 & 0.702 \\
\hline
\end{tabular}

\label{table:performance_difficulty}
\end{table}

\section{Conclusion and Future Work} \label{s:Conclusion}

In this work, we have: i) reviewed adaptive learning models and datasets; ii) introduced a VR platform for immersive education, specifically designed for adaptive learning, which monitors both biometric data and learning metadata; iii) released a public VR-based dataset for research in this area; iv) proposed five research challenges tied to this dataset; and v) designed two architectures for understanding detection to support dataset validation.


We believe that VR will transform education by improving presence, interactivity, and biometric monitoring compared to traditional e-learning. SMARTe-VR provides a comprehensive VR framework equipped with tools such as Auto QA, textbook creation, and biometric monitoring, improving educator capabilities and enabling AI-driven advancements. The SMARTe-VR dataset and its associated challenges offer a valuable resource for advancing AL models in VR, supporting research in IRT, KT, and MMLA.



SMART models offer an approach based on state-of-the-art IRT models, with a focus on understanding detection during learning. Using facial biometrics, they improve adaptability, enable real-time adjustments based on student states, and reduce the reliance on response data. These methods validated our data set with precision 85\%, although more research is needed to ensure a wider applicability.

In future work, we plan to expand the dataset with additional biometric modalities \cite{2018_INFFUS_MCSreview2_Fierrez}, more users, more questions per video, a broader range of lecture videos, and additional integrations of sensors and information sources \cite{2020_CDS_HCIsmart_Acien,pena23mm} to create a more robust resource. We also plan to monitor students in traditional e-learning settings under similar conditions to enable comparison between VR and traditional setups. 

In parallel, we will address the security, privacy and vulnerability aspects of VR-based and avatar-mediated learning environments \cite{acien20bots,pedrouzo2025isitreallyyou}. As these systems become increasingly immersive and realistic, it will be critical to ensure robust identity management through biometric and behavioral analysis \cite{kobayashi2019lifestyle,expressionauth}. The growing scientific interest in the security of avatar-based systems (exemplified by the IJCB 2025 Special Session on Tackling Safety Issues in the Emerging Cybernetic Avatar Era) underscores the need for secure biometric authentication and comprehensive safety frameworks in immersive environments. Addressing these challenges will be an integral part of our future research to further strengthen the security and robustness of the SMARTe-VR platform.

Furthermore, MR will be integrated into the platform to improve user experience. To address the challenges, we will share models with the community and refine the SMART-MLP and SMART-TCN models to detect understanding and include attention and fatigue detection. Ultimately, our goal is to incorporate these models into SMARTe-VR, creating a comprehensive adaptive learning experience that leverages the potential of VR-based education. 

Finally, this work has been strongly focused on e-learning applications, but the developed technologies also find a good fit in many other application areas like in VR-based entertainment, simulation, and real-time operation of simulators and certain equipment (e.g., medical or industrial inspection).

\begin{acks}
This work was supported by the following projects: BBforTAI (PID2021-127641OB-I00 MICINN/FEDER), HumanCAIC (TED2021-131787B-I00 MICINN), BIO-PROCTORING (GNOSS Program, Agreement MINDEF-UAM-FUAM dated 29-03-2022), M2RAI (PID2024-160053OB-I00 MICIU/FEDER), and ELLIS Unit Madrid. A. Morales is also supported by the Madrid Government in the line of Excellence for University Teaching Staff (V PRICIT). This work was conducted during the doctoral research stay of Roberto Daza at the Nagao Laboratory, Nagoya University. The authors are particularly grateful to the experimenters from Nagoya University and Hiroshima City University for their significant contributions to data collection and experiment implementation. In addition, we extend our thanks to both universities for providing experimental equipment and technical support, which were indispensable for the successful completion of this research. Without their support, this study would not have been possible.
\end{acks}

\bibliographystyle{ACM-Reference-Format}
\balance
\bibliography{sample-base}


\end{document}